\documentclass[12pt]{article}
\title{1/8 BPS black hole composites}
\usepackage{mathrsfs}
\usepackage{amsmath}

\global\arraycolsep=1pt
\usepackage[colorlinks=true,backref=true,linkcolor=black,anchorcolor=black,citecolor=black,filecolor   =black,menucolor=black,pagecolor=black,urlcolor=black]{hyperref}
\usepackage[Symbol]{upgreek}
\def\bea{\begin{eqnarray}}
\def\eea{\end{eqnarray}}
\def\be{\begin{equation}}
\def\ee{\end{equation}}{
\def\bem{\begin{multline}}
\def\eem{\end{multline}}

\usepackage{bbm}
\usepackage{dsfont}
\usepackage{amssymb}
\usepackage{textcomp}
\usepackage{wasysym}

\newcommand{\Scal}[1]{\Bigl ({#1} \Bigr )}
\newcommand{\scal}[1]{\bigl ({#1} \bigr )}
\newcommand{\CR}{\nonumber \\*}

\newcommand{\trace}{\hbox {Tr}~}

\DeclareMathAlphabet{\mathpzc}{OT1}{pzc}{m}{it}

\def\tt{{ \scriptscriptstyle t\, -1}}

\DeclareMathOperator{\ad}{ad}
\DeclareMathOperator{\Ad}{Ad}

\newcommand{\ord}[1]{{\scriptscriptstyle (#1)}}

\DeclareMathAlphabet{\mathpzc}{OT1}{pzc}{m}{it}

\renewcommand{\Im}{{\rm Im}}
\renewcommand{\Re}{{\rm Re}}
\newcommand{\Pfaff}{{\rm Pfaff}}

\usepackage[Symbol]{upgreek}
\usepackage{caption}
\usepackage{bbm}
\usepackage{graphicx}

\newcommand{\IR}{\mathbb{R}}
\newcommand{\IC}{\mathbb{C}}

\def\zero{{\mathpzc{0}}}
\def\un{{\mathpzc{1}}}
\def\deux{{\mathpzc{2}}}
\def\trois{{\mathpzc{3}}}

\def\n{{\mathpzc{n}}}

\newcommand{\sfrac}[2]{{\scriptstyle \frac{#1}{#2}}}

\def\ie{{\it i.e.}\ }
\def\eg{{\it e.g.}\ }

\def\pA{{\scriptscriptstyle A}}
\def\pB{{\scriptscriptstyle B}}

\def\e{\boldsymbol{e}}

\def\asym{{\scriptscriptstyle 0}}
\def\invo{{\APLstar}}

\def\C{{\bf Q}}

\def\P{\mathfrak{P}}

\def\V{{\mathcal{V}}}

\def\cM{\mathcal{M}}
\def\SU{SU_{\scriptscriptstyle \rm c}(8)}

\def\Ha{\mathcal{H}}
\def\Ka{\mathcal{K}}

\def\SpinS{Spin^*_{\scriptscriptstyle \rm c}}

\def\gl{\mathfrak{gl}}
\def\sl{\mathfrak{sl}}
\def\so{\mathfrak{so}}
\def\su{\mathfrak{su}}

\def\e{\mathfrak{e}}

\def\DJo{$\;$\kern-.4em \hbox{D\kern-.8em\raise.15ex\hbox{--}\kern.35em okovi\'c}}

\def\N{\mathcal{N}}
\newcommand{\eprint}[1]{{\href{http://arxiv.org/abs/#1}{\texttt{[#1}]}}}
\newcommand{\eprintN}[1]{{\href{http://arxiv.org/abs/#1}{\texttt{#1 [hep-th]}}}}

\begin{document}
\allowdisplaybreaks[1]
\renewcommand{\thefootnote}{\fnsymbol{footnote}}
\def\corr{$\spadesuit $}
\def\trefle{ $\clubsuit$}
\begin{titlepage}
\begin{flushright}
\
\vskip -3cm
{ \small AEI-2010-007}
\vskip 3cm
\end{flushright}
\begin{center}
{{\Large \bf
1/8 BPS black hole composites  }}
\lineskip .75em
\vskip 3em
\normalsize
{\large  Guillaume {\sc Bossard}\footnote{email address: bossard@aei.mpg.de}\\
\vskip 1 em
$^{*}$ {\it Max-Planck-Institut f\"{u}r Gravitationsphysik\\ Albert-Einstein-Institut\\ Am M\"{u}hlenberg 1, D-14476 Potsdam, Germany}\\
}

\vskip 1 em
\end{center}
\vskip 1 em
\begin{abstract}
We show that the 1/8 BPS condition for composite stationary black holes can be rewritten as a first order system of differential equations associated to the nilpotent orbit in which lie the Noether charges of the  black holes. Solving these equations, we prove that the most general 1/8 BPS black hole composites are solutions of the $\N=2$ truncation of the theory associated to the quaternions. This system of first order differential equations generalises to the non-BPS solutions with a vanishing central charge at the horizon in $\N=2,\, 4$ supergravity theories with a symmetric moduli space. We solve these equations for the exceptional $\N=2$ supergravity associated to the octonions. 
\end{abstract}

\end{titlepage}
\renewcommand{\thefootnote}{\arabic{footnote}}
\setcounter{footnote}{0}


\section{Introduction}

Exact solutions of supergravity theories defining effective theories of string theory compactifications on a  six-dimensional manifold are of prime importance for the understanding of non-perturbative properties of string theory \cite{PiolineLecture}. In particular, the derivation of BPS multi-black hole solutions within $\N=2$ supergravity defining bound states of the theory has permitted to explain the mismatch between the enumeration of spherically symmetric BPS black holes and the one of BPS states within the fundamental theory  \cite{Denef,Bates}. When considering compactifications on a $6$-torus, or $K3 \times T^2$, or orbifolds of them, it is possible to interpret certain solutions of the effective theory quantum mechanically by mean of string theory  computations at weak coupling. In $\N=4$ supergravity, the exact degeneracy of 1/4 BPS dyons is determined by an index formula  \cite{VVD,StroVD,JatkarSen,Dabholkar}, which large charges behaviour reproduces the classical Bekenstein--Hawking entropy of the corresponding black holes, providing in this way a statistical understanding of their thermodynamic. The validity of weakly coupled string theory descriptions of black hole micro-states at strong coupling is ensured by supersymmetry. It has been argued \cite{Dimitru,DabhoSen} that this non-renormalisation property would be a consequence of the attractor behaviour of extremal black holes \cite{FerraraAttra1, FerraraAttra2, FerraraAttra3, SenAttra,GoldsteinAttra}, and would thus apply more generally to (non-necessarily supersymmetric) extremal black holes. 

The most general BPS solutions are known in general \cite{Denef,Bates,Ortin}, although the absence of more general BPS solutions is still conjectural for maximal supergravity \cite{FerraraMagic}. For supergravity theories with scalar fields parameterising a symmetric space, the spherically symmetric extremal, but not BPS, solutions are by now pretty well understood  \cite{Ceresole:2007wx,Andrianopoli:2007gt,
Lopes Cardoso:2007ky,Perz:2008kh,Hotta:2007wz,Gimon:2007mh,Bellucci:2008sv,Gimon:2009gk,Gaiotto:2007ag,Ceresole:2009iy,Ferrara:2009bw}. In particular, the `fake superpotential' which determines the radial evolution of the scalar fields has been obtained in general in \cite{BossardW,Ceresole:2009vp}. For composite solutions including several non-BPS black holes, the static Papapetrou--Majumdar type solutions are also well understood \cite{Gaiotto:2007ag,NicolaiPM,BossardPM}. However, the much more interesting non-static bound states solutions carrying intrinsic  angular momentum have only been recently obtained for some particular axisymmetric examples within the $STU$ truncation of maximal supergravity \cite{Bena1,Bena2}. If the generalisation of these solutions to more general non-axisymmetric  configurations seems possible, it is technically very difficult. 

The success in obtaining the most general BPS solutions within the $\N=2$ supergravity theories is due to the existence of a first order system of differential equations, corresponding to the preservation of one-half of the supersymmetry charges, which solves the coupled Einstein equations \cite{Denef}. For spherically symmetric solutions, this first order system of differential equations is determined by the `fake superpotential'. Some attempts have been made in generalising the Denef construction to non-BPS configurations \cite{Gaiotto:2007ag,Galli}, but only static non-BPS solutions have been obtained. In this paper, we explain that a natural generalisation of the BPS first order differential equations is associated to extremal solutions in the theories admitting scalar fields parameterising a symmetric space, which results from the correspondence between extremal solutions and nilpotent orbit in Lie algebras.\footnote{See \eg \cite{Collingwood} for a thorough introduction to nilpotent orbits.}

The relation between BPS spherically symmetric black holes and nilpotent orbits has been first uncovered in \cite{Gunaydin:2005mx}. The precise  correspondence between the moduli space of spherically symmetric extremal black holes and Lagrangian subspaces of associated nilpotent orbits has been explained in general in \cite{BossardNil}. In \cite{NicolaiPM,BossardPM}, this relation has been used to characterise the static composite solutions of Papapetrou--Majumdar type, by mean of the `starred' Cayley triplet associated to nilpotent Noether charges. The same semi-simple element ${\bf h}$ of this `starred' Cayley triplet turns out to characterise the coset component of the Maurer--Cartan form ${\bf P}$ in the spherically symmetric case; and the `fake superpotential' has been derived using the first order equation $ [ {\bf h} , {\bf P} ]Ê= 2 {\bf P}$ in \cite{BossardW}. 

In this paper, we will show that this equation is still valid for non-static stationary BPS solutions, and is in fact equivalent to the preservation of 4 supercharges in solutions admitting only point-like singularities. The semi-simple generator ${\bf h}$ generalises the function $\alpha$ introduced by Denef in order to solve the equations of motion in $\N=2$ supergravity theories \cite{Denef}. 

We will first obtain this equation for 1/8 BPS solutions in maximal supergravity. Solving it, we will be able to prove that the most general 1/8 BPS black hole composites are necessarily defined in a maximal $\N=2$ truncation of the theory, as conjectured in \cite{FerraraMagic}. 

This first order system of differential equations can be generalised to non-BPS black hole solutions with vanishing central charges at the horizon in non-maximal supergravity theories. As a general example, we write and solve this system in the case of the exceptional $\N=2$ supergravity theory. The only essential data being then the symmetric tensor $c_{ijk}$ satisfying the adjoint identity which defines the cubic prepotential, the result extends straightforwardly to any $\N=2$ supergravity with a very special geometry.

Nevertheless, these solutions are rather trivial generalisations of the BPS solutions, and this paper is rather a necessary preliminary step toward the derivation of a system of first order differential equations for the non-BPS solutions with a non-vanishing central charge at the horizon. Such a system can be obtained from the parameterisation of the corresponding nilpotent orbit Lagrangian submanifold; but the determination of regular non-static stationary composite solutions solving it is out of the scope of this paper, and will only be discuss in a future publication.

\section{$E_{8(8)}$ invariant stationary equations}

We will first recall some properties of $\N=8$ supergravity in four dimensions in order to set up our notations, which are the same as in \cite{BossardW}. The massless scalar fields take values in  the symmetric space \cite{CremmerJulia}
\be 
\label{M4N8}
\cM_4 \cong \SU \backslash E_{7(7)} \ ,
\ee
where $\SU$ is the quotient of $SU(8)$ by the $\mathds{Z}_2$ centre leaving invariant the representations of even rank. According to the conventions of \cite{de Wit:1982ig} (up to normalisation factors), we write the coset representative $v$ as
\be v \, \Hat{=} \,  \left(\begin{array}{cc} {u_{ij}}^{IJ} & v_{ijKL} \\ v^{klIJ} & {u^{kl}}_{KL} \end{array}\right) 
\ ,
\ee
where little Latin letters are associated to the $SU(8)$ gauge symmetry, whereas capital Latin letters refer to the global $\SU \subset E_{7(7)}$. They both run from $1$ to $8$, and raising or lowering indices corresponds to complex conjugation (\eg $\Phi^{IJ} = ( \Phi_{IJ} )^*$ and $Z^{ij} = ( Z_{ij} )^*$). 
The invariant metric on $\cM_4$ can be written as 
\be 
ds^2_{\cM_4} =  \frac{1}{24} V_{ijkl} V^{ijkl} \ ,
\ee
where 
\be 
V_{ijkl} ={u_{ij}}{}^{IJ} d v_{klIJ} - v_{ijIJ} d {u_{kl}}{}^{IJ}  
 \label{Evielbeins} 
\ee
are  the $\SU \backslash E_{7(7)}$ vielbeins, which define automatically a complex self-dual antisymmetric tensor by property of the $\e_{7(7)}$ Lie algebra.

As explained in \cite{Breitenlohner:1987dg}, the dimensional 
reduction of $\N=8$ supergravity along the time direction leads
to a non-linear sigma model on  
\be 
 \cM_3^* \cong \SpinS(16) \backslash E_{8(8)} \ ,
 \ee
where $\SpinS(16)$ is the quotient of $Spin^*(16)$ by the $\mathds{Z}_2$ subgroup that acts trivially in the chiral Weyl representation. To parameterise this space in a way suited to the dimensional reduction, 
recall that the Lie algebra $\e_{8(8)}$ admits the real five-graded decomposition 
\be 
\e_{8(8)}\cong {\bf 1}^{\ord{-2}} 
\oplus {\bf 56}^{\ord{-1}} \oplus \scal{\gl_1 \oplus 
\e_{7(7)}}^\ord{0} \oplus {\bf 56}^\ord{1} 
\oplus {\bf 1}^\ord{2} \ ,
\label{five}
\ee
such that $\e_{7(7)}$ is the Lie algebra of the four-dimensional duality group, and $\sl_2 \cong  {\bf 1}^{\ord{-2}} \oplus {\gl_1}^\ord{0} \oplus {\bf 1}^\ord{2}$ the Lie algebra of the Ehlers duality group for stationary solutions. We write the generators of $\e_{7(7)} \cong \su(8) \oplus {\bf 70}$ as ${{\bf G}_I}^J,\, {\bf G}_{IJKL}$ and the ones of $\sl_2 \cong  {\bf 1}^{\ord{-2}} \oplus {\gl_1}^\ord{0} \oplus {\bf 1}^\ord{2}$ as ${\bf F},\, {\bf H} ,\, {\bf E}$, respectively. The generators of grade $1$ and $-1$ will be written as ${\bf E}_{IJ},\, {\bf E}^{IJ}$ and ${\bf F}_{IJ},\, {\bf F}^{IJ}$, such that they only appear in $\e_{8(8)}$ through the combinations 
\be X_{IJ} {\bf E}^{IJ} - X^{IJ} {\bf E}_{IJ} \ ,  \hspace{10mm}Y_{IJ} {\bf F}^{IJ} - Y^{IJ} {\bf F}_{IJ} \, .\ee

The negative weight part of the the five-graded decomposition (\ref{five})
 \be 
 \mathfrak{p} \cong  {\bf 1}^\ord{-2} \oplus  {\bf 56}^\ord{-1}  \oplus  \scal{\gl_1 \oplus \e_{7(7)}}^\ord{0}   
\label{parabolic}
\ee
defines the Lie algebra of a maximal parabolic subgroup $\P\subset E_{8(8)}$. $\SU \backslash \P$ 
is isomorphic to the Riemannian symmetric space $\cM_3 \cong Spin_{\scriptscriptstyle \rm c}(16)\backslash E_{8(8)}$ by the Iwasawa decomposition, and to a dense subset of 
the pseudo-Riemannian symmetric space $\cM_3^*$. A generic 
element of $\SU \backslash\P$ may be parameterised as
\be 
\V = \Ad(v)\, \exp\left(  U \,   {\bf H}\right) \,  \exp\Scal{\sigma\, {\bf F} + 
\Phi_{IJ} {\bf F}^{IJ} - \Phi^{IJ} {\bf F}_{IJ}  } \ .
 \label{Pgauge} 
\ee
$U$ is identified as the scale factor in the metric Ansatz
\be ds^2 = - e^{2U} \scal{dt + \omega_\mu dx^\mu }^2 + e^{-2U} \gamma_{\mu\nu} dx^\mu dx^\nu \ ,  \ee
and $\sigma$ is defined from the Kaluza--Klein vector $\omega_\mu$ via its equation of motion
\be d \sigma = - e^{4U} \star d \omega - \frac{i}{2} \scal{ \Phi^{IJ} d \Phi_{IJ} - \Phi_{IJ} d \Phi^{IJ} }\, , \ee
where $\star$ is the Hodge star operator on the three-dimensional Riemannian base space equipped with the metric $\gamma_{\mu\nu}$.\footnote{Strictly speaking, $\gamma_{\mu\nu}$ is only Riemannian outside the horizons, and is in fact degenerated at the horizon of a non-extremal black hole.}  $v$ is the coset representative in \eqref{M4N8}, and $\Phi_{IJ}$ are the duality covariant scalars associated to the electromagnetic fields  transforming 
as an antisymmetric complex tensor of $\SU \subset E_{7(7)}$. The electromagnetic fields can be recovered by decomposing $\Phi_{IJ}$ in a first step according to a Darboux basis associated to the $E_{7(7)}$ symplectic form, as
\be \Phi_{IJ} = \xi_{IJ} + i \tilde{\xi}_{IJ} \ , \ee
and then dualising the 28 real fields $\tilde{\xi}_{IJ}$ according to their equations of motion 
\bem d \Hat{\xi}_{IJ} = - e^{-2U}\star \Bigl( d \tilde{\xi}_{IJ} + 2 \Re \bigl[v_{ijIJ} \scal{u^{ij}{}_{KL} + v^{ijKL} }\bigr]d \tilde{\xi}_{KL}  \Bigr . \\* \Bigl . - 2 \Im \bigl[v_{ijIJ} \scal{u^{ij}{}_{KL} - v^{ijKL} } \bigr]d \xi_{KL} \Bigr) -   \xi_{IJ}  d \omega  \ , 
\end{multline}
such that the 28 vector fields are 
 \be A_{IJ} = \xi_{IJ} \scal{dt + \omega }+ \Hat{\xi}_{IJ}\, . \ee
The associated Maurer--Cartan form decomposes into its coset 
and $\so^*(16)$ components according to
\be
d \V  \, \V^{-1}  = {\bf B} +{\bf P}  \quad, \qquad {\bf B} \in \so^*(16) 
  \;\; , \;\;\; 
 {\bf P} \in \e_{8(8)} \ominus\so^*(16) \, .
\ee
A straightforward computation gives
\begin{multline}  
\label{P}
{\bf P} =  dU   \, {\bf H}  + \frac{1}{2} e^{-2U} \Scal{  d\sigma + \frac{i}{2} \scal{ \Phi^{IJ}  d{ \Phi}_{IJ} - \Phi_{IJ}  d{ \Phi}^{IJ}   } } ( {\bf F} + {\bf E})  \\* + \frac{1}{2}e^{-U}  \,\Scal{   \scal{{u_{ij}}^{IJ} d{\Phi}_{IJ} - v_{ijIJ} d{\Phi}^{IJ} } \scal{{\bf F}^{ij} - {\bf E}^{ij} } -  \scal{{u^{ij}}_{IJ} d{\Phi}^{IJ} - v^{ijIJ} d{\Phi}_{IJ} } \scal{{\bf F}_{ij} - {\bf E}_{ij} } } \\* + \frac{1}{24}\Scal{{u_{ij}}{}^{IJ} d{v}_{klIJ} - v_{ijIJ} {d{u}_{kl}}{}^{IJ}}   {\bf G}^{ijkl}   \ ,\end{multline}
where the $\e_{8(8)}$ generators with lowercase indices $i,\, j,\, \cdots$ satisfy
the same commutations rules as the ones with capital indices.

The equations of motion  then take the manifestly $E_{8(8)}$ invariant form
\be
R_{\mu\nu} = \trace {\bf P}_\mu {\bf P}_\nu  \ ,\qquad d \star  \,\Scal{ \V^{-1}{\bf P} \V } = 0 \, ,
\label{EinsteinE}
\ee
where the trace is normalised such that $ \trace {\bf H}^2 = 2$. The solutions admitting only particle-like sources coupled to the electromagnetic fields, \ie multi-black hole solutions, are instantons within the three-dimensional theory. To each black hole corresponds a $2$-cycle $\Sigma$ in the three-dimensional base space, and the associated $\e_{8(8)}$-valued Noether charge 
\be 
\C_{|\Sigma}  \equiv  \frac{1}{4\pi } \int_\Sigma \V^{-1}{\bf P} \,  \V\ . \label{Noether}
\ee
The absence of naked singularities requires that it satisfies the characteristic equation
\be {\C_{|\Sigma}}^5 - \frac{5}{2} \trace {\C_{|\Sigma}}^2\,  \cdot {\C_{|\Sigma}}^3 + \trace^{\hspace{-1mm} 2 \hspace{1mm}}   {\C_{|\Sigma}}^2 \, \cdot  {\C_{|\Sigma}} = 0 \label{chara} \ , \ee
in the $\bf 3875$ irreducible representation of $E_{8(8)}$ that appears in the symmetric tensor product of two copies of the adjoint representation \cite{BossardNil}. In general, the existence of other black holes modify the expression of the horizon area $A_{|\Sigma}$ and the surface gravity $\kappa_{|\Sigma}$ of a given black hole as a function of its charges, as can be exhibited in axisymmetric multi-NUT solutions \cite{BossardNUT}. Nonetheless, for black holes admitting no intrinsic angular momentum, their product is still entirely characterised by the charge as 
\be A_{|\Sigma} \, \, \kappa_{|\Sigma} = 4 \pi  \sqrt{  \frac{1}{2}Ê\trace {\C_{|\Sigma}}^2 } \, .\ee
It follows that for extremal black holes, $ \trace {\C_{|\Sigma}}^2 = 0 $, and so from (\ref{chara}) that the Noether charge $\C_{|\Sigma}$ is nilpotent of order five in the ${\bf 3875}$ of $E_{8(8)}$,
\be {\C_{|\Sigma}}^5 = 0  \, .\ee
Regular black holes with a non-vanishing horizon area are associated to generic Noether charges satisfying this condition, for which 
\be  \ad_{{\C_{|\Sigma}}}^4 \ne 0\ ,  \ee
in the adjoint representation. Such charges lie in one of the two nilpotent orbits of $E_{8(8)}$ of dimension 114 \cite{BossardNil}. These two nilpotent orbits can be distinguished by the semi-simple component of the stabiliser of $\C_{|\Sigma}$ inside $E_{8(8)}$, which is homomorphic to the stabiliser of the corresponding electromagnetic charges $Q_{IJ}$ inside $E_{7(7)}$ \cite{BossardW}, \ie either $E_{6(2)}$ for 1/8 BPS black holes or $E_{6(6)}$ for non-BPS black holes \cite{Ferrara:1997uz,Bellucci:2006xz}.

Several 1/8 BPS black holes can be in equilibrium, such that they define a regular stationary solution of Einstein equations \cite{FerraraMagic}. The same will be true for several non-BPS black black holes \cite{Bena1,Bena2}. However, the existence of such solutions is associated to the fact that the moduli $v$ permit to interpolate between the electromagnetic charges $Q_{IJ}$ of the different black holes defining the composite. So it is rather clear that two black holes of different type (one 1/8 BPS and the other not) cannot be in equilibrium. 

\section{1/8 BPS first order system}

It is convenient to decompose $Spin^*(16)$ in terms of its maximal compact subgroup $U(8)$. Considering a basis of fermionic oscillators to define the spinor representation, the coset $1$-form $P$ can be written \cite{BossardW,BossardNil}
\be
 | P \rangle = ( 1 + \invo ) \Bigl( d U - \frac{i}{2} e^{2U} \star d \omega  + e^{-U} \, v^\tt(d\Phi)_{ij} \, a^i a^j  \Bigr . \\* \Bigl .  + \frac{1}{24} V_{ijkl} \,  a^i a^j a^k a^l  \Bigr) | 0 \rangle \ , 
 \label{Pspinor} \ee
where $\invo$ is the anti-involution of $Spin^*(16)$ that defines the Majorana--Weyl reality condition, which acts as an $SU(8)$ hodge star operator, and 
\be  v^\tt(d\Phi)_{ij} \equiv u_{ij}{}^{IJ} d \Phi_{IJ} - v_{ijIJ} d \Phi^{IJ} \ . \ee

The bosonic component of the supersymmetry transformation of the three-dimensional gravitino field is
\be \delta \psi^i_\alpha = \nabla_{\hspace{-0.5mm} B}  \, \epsilon^i_\alpha \label{susyP} \ , \ee
and the one of the dilatino field is \cite{BossardNil}
\be
\delta \left| \chi  \right>_\alpha  = e^\mu_a    {{\sigma^a }_\alpha}^\beta 
  \, \Scal{   \epsilon^i_\beta \, a_i +  \varepsilon_{\beta\gamma}    
  \epsilon_i^\gamma   a^i  } \left| P_\mu \right> \, .   \label{ChiVar} 
 \ee
The leading component of ${\bf P}$ in the asymptotic region of a black hole composite solution is spherically symmetric 
\be {\bf P} \sim - \V_\asym \left(Ê\sum_\pA \C_{\pA}\right)  \V^{-1}_{\asym} \, d \, \frac{1}{r} \ , \label{Sphe} \ee
the angular momentum contribution being subleading. In the neighbourhood of a black horizon, it follows from (\ref{Noether}) that ${\bf P}$ is approximately given by
\be {\bf P} \sim - \V(x_\pA)   {\bf Q}_{\pA} \V^{-1}(x_\pA)  \, d \, \frac{1}{|x-x_\pA|} \ . \ee
We conclude that the right-hand-side of (\ref{ChiVar}) factorises in these regions, such that by continuity, its vanishing implies
\be  \Scal{   \epsilon^i_\alpha \, a_i +  \varepsilon_{\alpha\beta}    
  \epsilon_i^\beta   a^i  } \left| P_\mu \right> = 0 \ . \label{susyC}
\ee
This is the 1/8 BPS condition for solutions involving only point-like sources (as opposed to solutions involving string-like sources \cite{Ortin}). It implies 
\be R_{\mu\nu} =  \langle P_\mu | P_\nu \rangle = 0 \ , \ee
such that $\gamma_{\mu\nu}$ is the Euclidean metric $\delta_{\mu\nu}$ on $\IR^3$. 
Its validity for four spinor parameters $\epsilon^i_\alpha$ moreover implies that the latter satisfy the symplectic Majorana condition  
\be  \epsilon_i^\alpha + \varepsilon^{\alpha\beta} \Omega_{ij} \epsilon^j_\beta = 0\ ,   \label{Maj} \ee
for a rank two antisymmetric tensor $\Omega_{ij} \in SU(8) / ( SU(2) \times SU(6))$ satisfying 
\be \Omega_{[ij} \Omega_{kl]} = 0 \ ,\qquad \Omega^{ij} \Omega_{ij} = 2 \label{BPSattra}\ ,  \ee
such that $I_i^j  \equiv \Omega_{ik} \Omega^{jk} $ is a projector onto a $\IC^2 \subset \IC^8$ subspace, and
\be I_i^j \epsilon_j^\alpha = \epsilon_i^\alpha \, . \label{Maji} \ee
Using (\ref{Maj}) and (\ref{Maji}) in (\ref{susyC}), one then obtains that 
\be \Omega^{ik}  e^{-U}  v^\tt(d\Phi)_{jk}  = I^i_j \Scal{\frac{1}{2} dU - \frac{i}{4} e^{2U} \star d \omega } \ , \label{ConPhi} \ee
such that the vanishing of (\ref{susyP}) simplifies to 
\be d_B \epsilon^i_\alpha = D \epsilon^i_\alpha  + \Scal{\frac{1}{2}d U + \frac{i}{4} e^{2U} \star d \omega }\epsilon^i_\alpha = 0 \label{PreK} \ , \ee
where $D$ is the $SU(8)$ covariant derivative 
\be D \epsilon^i_\alpha = d \epsilon^i_\alpha - \frac{1}{3} \scal{u_{jk}{}^{IJ} d u^{ik}{}_{IJ} - v_{jkIJ} d v^{ikIJ }} \epsilon^j_\alpha \ .\ee
Using the Bianchi identity 
\be d {\bf B} + {\bf B}^2 + {\bf P}^2 = 0 \ ,  \ee
and the fact that (\ref{susyC}) implies that ${\bf P}^2$ leaves $\epsilon^i_\alpha$ invariant as an $\so^*(16)$ generator, we conclude that the Killing equation (\ref{PreK}) is integrable,
\be {d_B}^2\,  \epsilon^i_\alpha = 0 \ . \ee
The compatibility of the Killing spinor equation (\ref{PreK}) and the reality condition (\ref{Maj}) implies that the tensor $\Omega_{ij}$ satisfies the differential equation 
\be D \Omega_{ij} + \frac{i}{2} e^{2U}\star d \omega \, \Omega_{ij} = 0 \label{Killing} \ .\ee 
This equation together with 
\be \scal{I^i_j a^j + \Omega^{ij} a_j } |P \rangle = 0\ ,  \label{BPSC} \ee
defines a system of first order differential equations that solve the Einstein equations (\ref{EinsteinE}) and implies that the corresponding solutions are 1/8 BPS. 

The aim of this paper is to understand this first order system of differential equations independently of supersymmetry, in terms of a `stared' Cayley triplet associated to the nilpotent orbit in which lie the Noether charges $\C_{|\Sigma}$.

Using the identity 
\be \Omega_{ij} a^i a^j - \Omega^{ij} a_i a_j = 2 + \Omega_{ij} a^i \scal{a^j + \Omega^{jk} a_k } -  \Omega^{ij} a_i \scal{a_j - \Omega_{jk} a^k } \ , \ee
one obtains that the BPS condition (\ref{BPSC}) can be rewritten as
\be h| P \rangle = 2 |P \rangle \ ,  \label{triplet} \ee
for the $\so^*(16)$ generator 
\be h\equiv \Omega_{ij} a^i a^j - \Omega^{ij} a_i a_j  \ee
satisfying (\ref{Killing}). 

It follows that for 1/8 BPS solutions, not only the Noether charge $\C_{|\Sigma}$ lies inside the nilpotent orbit of $E_{8(8)}$ characterised by the semi-simple stabiliser $E_{6(2)}$, but the $1$-form ${\bf P}$ itself lies inside the Lagrangian submanifold of this nilpotent orbit defined by its intersection with the coset component $\e_{8(8)} \ominus \so^*(16)$ 
\be | P \rangle \in \SpinS(16) \big /  {\scal{SU(2) \times SU(6)} \ltimes \scal{ ( { \bf 2} \otimes {\bf 6})^\ord{2}\oplus \mathds{R}^\ord{4}}} \ ,\ee
and in particular, $ {{\bf P}_{\mu}}^5 = 0 $ in the ${\bf 3875}$ representation for all $\mu = 1, \, 2 ,\, 3$. 

Written in this way, the first order system can be defined independently of supersymmetry as the requirement that ${\bf P}$ belongs to the intersection of the nilpotent orbit of $E_{8(8)}$ with the coset component $\e_{8(8)} \ominus \so^*(16)$ defined by the generator ${\bf h}$. Equation (\ref{Killing}) can then be reinterpreted as a consistency condition for $h | P \rangle = 2 |P \rangle$. Writing $B$ in terms of fermionic oscillators as
\bem B = \frac{i}{4} \star d \omega [ a^i , a_i ]+ \frac{1}{2} e^{-U} \Scal{ v^\tt(d \Phi)_{ij} \, a^i a^j - v^\tt(d\Phi)^{ij} \, a_i a_j } \\* + \frac{1}{6}\scal{u_{jk}{}^{IJ} d u^{ik}{}_{IJ} - v_{jkIJ} d v^{ikIJ }}  [a^j , a_i ]\ ,\end{multline}
one has the consistency conditions 
\be   \scal{ d h - [B , h ] }{}_{\, \wedge} | P \rangle = 0 \,, \qquad   \scal{ d h - [B , h ] } \star  | P \rangle = 0 \ ,\label{Consi} \ee
that follow form the equation of motion and the Bianchi identity, respectively. Using (\ref{ConPhi}), one obtains that 
\bem d h - [ B , h ]=  \Scal{D \Omega_{ij} + \frac{i}{2} e^{2U} \star d \omega \, \Omega_{ij} }  a^i a^j -  \Scal{D \Omega^{ij} - \frac{i}{2} e^{2U} \star d \omega \, \Omega^{ij} }   a_i a_j \\* - \frac{i}{2} e^{2U} \star d \omega \, \Omega^{ij} \scal{a_i + \Omega_{ik} a^k } \scal{a_j + \Omega_{jl} a^l } \ ,\end{multline} 
where the last term is a generator of grade $4$ with respect to $h$,
\be \bigr[ h , \Omega^{ij} \scal{a_i + \Omega_{ik} a^k } \scal{a_j + \Omega_{jl} a^l } \bigl]= 4 \, \Omega^{ij} \scal{a_i + \Omega_{ik} a^k } \scal{a_j + \Omega_{jl} a^l } \ , \ee
and therefore does not contribute in (\ref{Consi}), such that one recovers (\ref{Killing}).

The advantage of this formulation is that it will be then easily generalised to non-BPS solutions with saturated charges in less supersymmetric theories, such as the non-BPS solutions of $\N=2$ supergravity with a vanishing central charge at the horizon, $Z_* = 0$. 

\section{The solution}

To solve this system of equations, let us consider in a first step the integrability condition 
\be  h^2 | P \rangle = 4 | P \rangle \ , \ee
to Equation (\ref{triplet}). It implies that 
\be  2 I_{[i}^k v^\tt(d \Phi)_{j]k}Ê+ \Omega_{ij} \Omega^{kl} v^\tt(d \Phi)_{kl}Ê= 0 \ , \qquad 2 I_{[i}^p \,  V_{jkl]p} + 3 \Omega_{[ij} \Omega^{pq}\, ÊV_{kl]pq} = 0 \ , \label{N2truncation} \ee
or in other words, that within the decompositions 
\be {\bf 28}Ê\cong \mathds{C}Ê\oplus {\bf 2} \otimes {\bf 6} \oplus {\bf 15} \ , \qquad {\bf 70}Ê\cong ({\bf 2}Ê\otimes {\bf 20})_\mathds{R} \oplus {\bf 15} \ , \ee
 associated to the stabiliser subgroup $SU(2) \times SU(6) \subset SU(8)$ of $\Omega_{ij}$, the component of $v^\tt(d \Phi)$ in the ${\bf 2} \otimes {\bf 6}$ and the component of $V_{ijkl}$ in the $({\bf 2}Ê\otimes {\bf 20})_\mathds{R} $ vanish.

Looking at equation (\ref{Killing}) one sees that the only way the projector $I_i^j \equiv \Omega_{ik} \Omega^{jk}$ can vary over space-time is if the $SU(8)$ connexion $ \frac{1}{3} u_{jk}{}^{IJ} d u^{ik}{}_{IJ} - v_{jkIJ} d v^{ikIJ }  $ admits a non-trivial component in the ${\bf 2} \otimes {\bf 6}$ of the stabiliser subgroup $SU(2) \times SU(6)$ of $\Omega_{ij}$. 

The solution is by hypotheses asymptotically spherically symmetric at leading order (\ref{Sphe}), and the corresponding three-dimensional geometry is flat. Defining spherical coordinates $r,\, \theta ,\, \varphi$, one can solve the equations of motion perturbatlively in $1/r$. At first order in $1/r$, the solution is identical to a spherically symmetric black hole solution, therefore $\Omega_{ij}$ is constant at this order and $v$ is an element of the $Spin^*(12)$ subgroup associated to the $\N=2$ truncation determined by $\Omega_{ij}$. It follows that at this order, the $SU(8)$ connexion is an element of $\mathfrak{u}(6)$, and only the phase of $\Omega_{ij}$ can be modified at second order in $1/r$. Using (\ref{N2truncation}), one concludes that the scalar fields still lie in the same $Spin^*(12)$ subgroup at second order in $1/r$. Recursively, one concludes that $I_i^j$ is a constant over space-time, and that the 1/8 BPS solutions of maximal supergravity are also 1/2 BPS solutions of the $\N=2$ truncation of the theory associated to $I_i^j$ and the asymptotic value of the scalar fields. 

This establishes the result argued in \cite{FerraraMagic}, \ie the most general 1/8 BPS multi-black hole solutions of maximal supergravity can be obtained by embedding the Denef solutions of the $\N=2$ supergravity associated to the magic square corresponding to the quaternions \cite{GunaydinMagic}, which bosonic sector is identical to the one of $\N=6$ supergravity. 

Let us discuss these solutions more explicitly within maximal supergravity. The direct generalisation of Denef Ansatz for the electromagnetic fields \cite{Denef} is 
\be \Phi_{IJ} = \frac{1}{2} e^{U} \scal{u^{ij}{}_{IJ} \Omega_{ij} + v_{ijIJ} \Omega^{ij} } \ , \label{Ansatz} \ee
in terms of which the equation of motion of $\Phi_{IJ}$ reduces to 
\bem d \Bigl[- \star d \Scal{\frac{1}{2} e^{-U} \scal{ u^{ij}{}_{IJ}  \Omega_{ij} - v_{ijIJ} \Omega^{ij} } } \Bigr . \\* \Bigl . + e^{-U}  \star  \Scal{u^{ij}{}_{IJ} \scal{D \Omega_{ij} +  \frac{i}{2} e^{2U}\star d \omega \, \Omega_{ij} } - v_{ijIJ} \scal{D \Omega^{ij} -  \frac{i}{2} e^{2U}\star d \omega \, \Omega^{ij} } } \Bigr] = 0 \ , \end{multline}
such that using (\ref{Killing}),
\be  d \star d \Scal{\,  e^{-U} \scal{ u^{ij}{}_{IJ}  \Omega_{ij} - v_{ijIJ} \Omega^{ij} } } = 0\ ,  \ee
and $\Omega_{ij}$ is defined in functions of, a priori, 56 harmonic functions $\Ha_{IJ}$ on $\IR^3$ as  
\be  \Omega_{ij}   = 2 e^{U} \scal{u_{ij}{}^{IJ}\Ha_{IJ} + v_{ijIJ} \Ha^{IJ} } \, .  \label{OmegaCons} \ee
However, Equation (\ref{N2truncation}) implies that the harmonic functions $\Ha^{IJ}$ must satisfy the 24 constraints 
 \be 2 I_{[i}^k v(d \Ha)_{j]k} + \Omega_{ij} \Omega^{kl} v(d\Ha)_{kl} = 0\ . \ee
which restrict the number of independent harmonic functions to the 32 functions of the corresponding $\N=2$ truncation. 

The existence of an $E_{7(7)}$ transformation relating $\Ha_{IJ}$ to a tensor $\Omega_{ij}$ satisfying  (\ref{BPSattra}) requires the quartic $E_{7(7)}$ invariant 
\be \lozenge(\Ha) =  16 \left( \Ha_{IJ} \Ha^{JK} \Ha_{KL} \Ha^{LI} - \frac14 \scal{\Ha_{IJ} \Ha^{IJ} }^2 + 4 
\Scal{\Pfaff(\Ha) + \overline{\Pfaff( \Ha)} } \right) > 0\ , 
\label{defloz}
\ee
to be strictly positive on $\IR^3$. In such a case, an attractor submanifold 
 \be \cM_{\scriptscriptstyle \rm BPS} \cong ( SU(2) \times SU(6) ) \backslash E_{6(2)} \subset \cM_4 \ee of $E_{7(7)}$ elements  satisfying this condition is guarantied to exist \cite{Bellucci:2006xz,Andrianopoli:1997wi}, which corresponds to the minimum of the 1/8 BPS fake superpotential $W = \Omega^{ij} v(\Ha)_{ij}$. The condition $ \frac{\partial W}{\partial v} = 0 $ can be written in terms of 
\be v(\Ha)_{ij} \equiv {u_{ij}{}^{IJ} \Ha_{IJ} + v_{ijIJ} \Ha^{IJ} } \ , \ee
as the algebraic equation
\be v(\Ha)_{[ij} v(\Ha)_{kl]} = 0 \ ,    \label{BPSattractor} \ee
which is manifestly consistent with (\ref{BPSattra}) and (\ref{OmegaCons}). Differentiating this equation, one obtains that
\be V_{ijkl} v(\Ha)^{kl} + v(d \Ha)_{ij} + 2 I_{[i}^k v(d \Ha)_{j]k} + \frac{1}{2} \Omega_{ij} \Omega^{kl} v(d\Ha)_{kl} = 0\ ,  \ee
and the normalisation condition implies 
 \be e^{-2U}= 2 v(\Ha)^{ij} v(\Ha)_{ij} = \sqrt{\lozenge(\Ha)} \ .\ee  
 Differentiating the latter expression with respect to the harmonic functions $\Ha_{IJ}$, one obtains
  \begin{gather} \Omega_{ij} u^{ij}{}_{IJ} = \frac{1}{2 \sqrt[4]{ \lozenge(\Ha)}} \Scal{Ê\frac{\partial \sqrt{ \lozenge(\Ha)}}{\partial \Ha^{IJ}} + 2 \Ha_{IJ}}  \ , \quad \Omega^{ij} v_{ijIJ} = \frac{1}{2 \sqrt[4]{ \lozenge(\Ha)}} \Scal{Ê\frac{\partial \sqrt{ \lozenge(\Ha)}}{\partial \Ha^{IJ}} - 2 \Ha_{IJ}}  \ , \label{scalaires} \CR
  \Phi_{IJ} = \frac{1}{2 \lozenge(\Ha)} \, \frac{ \partial \lozenge(\Ha)}{ \partial \Ha^{IJ}} \ , \end{gather}
which determines the scalar fields up to 40 functions parameterising the attractor manifold $\cM_{\scriptscriptstyle \rm BPS}$ associated to $\Ha_{IJ}$. Condition (\ref{N2truncation}) gives furthermore
\be V_{ijkl} = - \frac{12}{\sqrt{\lozenge(\Ha)}} \Scal{v(\Ha)_{[ij} v(d \Ha)_{kl]} + \frac{1}{24} \varepsilon_{ijklmnpq} v(\Ha)^{mn} v(d \Ha)^{pq} } \, , \ee
such that $V_{ijkl}$ is zero in the associated flat directions, and these 40 functions are determined in function of the value of the moduli $v$ in the asymptotic region. Note that $\Omega_{ij}$ varies only through its overall phase, which is itself fixed by choosing a gauge for the scalar fields.

One can now check that this solution is compatible with the Killing spinor equation (\ref{Killing}), 
  \bea 
   0 &=& D \scal{ e^{U} v(\Ha)_{ij} } + \frac{i}{2} e^{3U} \star  d \omega \, v(\Ha)_{ij} \CR
    &=& e^{U} \biggl(  \Scal{d U + \frac{i}{2} e^{2U} \star d \omega }v(\Ha)_{ij} + V_{ijkl} v(\Ha)^{kl} + v(d\Ha)_{ij} \biggr) \CR 
 & =& e^{3U} \Scal{ \frac{i}{2}\star d \omega + \Ha^{IJ} d \Ha_{IJ} - \Ha_{IJ} d \Ha^{IJ} } v(\Ha)_{ij}
\eea
where we used the $E_{7(7)}$ invariance of the symplectic form in the last line, \ie 
\be v(\Ha)^{ij} v(d\Ha)_{ij} - v(\Ha)_{ij} v(d\Ha)^{ij} =  \Ha^{IJ} d \Ha_{IJ} - \Ha_{IJ} d \Ha^{IJ} \, . \ee
This gives the first order equation
 \be d \omega = 2 i \star \Scal{  \Ha^{IJ} d \Ha_{IJ} - \Ha_{IJ} d \Ha^{IJ} }\, .\ee

In order to check the compatibility of this solution with equation (\ref{triplet}), it is convenient to use the Ansatz (\ref{Ansatz}), and the Killing spinor equation  (\ref{Killing}) to rewrite $|P\rangle$ as
\bem
 | P \rangle = ( 1 + \invo ) \Bigl( d U - \frac{i}{2} e^{2U} \star d \omega  - e^{U} \scal{v(d \Ha)_{ij} + i e^{2U}  \star d \omega \, v(\Ha)_{ij} } \, a^i a^j  \Bigr . \\* \Bigl .  + \frac{1}{24} V_{ijkl} a^i a^j a^k a^l  \Bigr) | 0 \rangle  \, .\end{multline}
One checks indeed that the solution satisfies the condition ${\bf h}| P \rangle = 2 | P \rangle$.  

The general solution is therefore defined in function of 32 harmonic functions 
\be \Ha_{IJ} = \sum_\pA   \frac{Q^\pA_{IJ}}{|x - x_\pA|} + \frac{1}{2} \scal{u_{\asym}^{\, ij}{}_{IJ} \Omega_{\asym ij} - v_{\asym \, ijIJ} \Omega_{\asym}^{ij} } \ , \ee
where the $Q^\pA_{IJ}$ are the electromagnetic charges of the black holes constituting the composite solution, $v_{\asym}$ is the asymptotic value of the moduli and $\Omega_{\asym \, ij}$ is the asymptotic tensor associated to the 4 asymptotically preserved supercharges. 

 Note that the problem of determining the allowed asymptotic value of the scalar fields for which the solution exists reduces to the problem of determining them within the $\N=2$ truncation. Indeed, any element of $E_{7(7)}$ can be written as the product of an element of $SU(8)_{\scriptscriptstyle \rm c}$, an element of $E_{6(2)}$  and an element of $Spin^*(12)$; and, both the harmonic functions and  the equations (\ref{scalaires}) determining the scalar fields (up to this constant value in $\cM_{\scriptscriptstyle \rm BPS}$), are invariant with respect to an $E_{6(2)}$ left multiplication of $v$ leaving invariant $\Omega_{\asym\, ij}$.\footnote{The $E_{7(7)}$ left multiplication is not a symmetry of the theory as opposed to the right multiplication, however the $E_{6(2)}$ left multiplication is a symmetry of the 1/8 BPS solutions.} It follows that only the $Spin^*(12)$ factor is constrained.
 
 $\Omega_{\asym \, ij}$ is itself determined in function of the asymptotic moduli $v_{\asym}$ and the electromagnetic charges such that 
\be Z_{\asym\, ij} \equiv  \sum_\pA \scal{ u_{\asym\, ij}{}^{IJ} Q^\pA_{IJ} + v_{\asym\, ijIJ} Q^{\pA\, IJ}} = \frac{1}{2} \Scal{M \, \Omega_{\asym\, ij} + \rho_\un \Omega^\un_{ij} + \rho_\deux \Omega^\deux_{ij} + \rho_\trois \Omega^\trois_{ij} }\ , \ee
 where $\Omega^\n_{ij}$ are orthonormal tensors of rank two defining with $\Omega_{\asym\, ij}$ the standard skew diagonalisation of the asymptotic central charge $Z_{\asym\, ij}$, such that $M$ is the ADM mass of the solution, which satisfies 
 \be M > \rho_\un \ge \rho_\deux \ge \rho_3 \ . \ee

In order to avoid Dirac--Misner string singularities, one must also make sure that the form $\omega$ is globally defined on $\IR^3$, such that 
\be 0 =  d d \omega  = 2i \scal{ \Ha^{IJ} d \star d \Ha_{IJ} - \Ha_{IJ} d \star d \Ha^{IJ} }\ , \ee
which requires the absence of NUT charge at each centre \cite{Denef}, \ie 
\be \sum_{\pB \ne \pA} \frac{ Q^{\pA\, IJ} Q^\pB_{IJ} - Q^\pA_{IJ} Q^{\pB\, IJ} }{ |x_\pA - x_\pB | } = \frac{1}{2} \Scal{\Omega^{ij}_\asym \,  v_\asym(Q^\pA)_{ij} -  \Omega_{\asym\, ij} \,  v_\asym(Q^\pA)^{ij} }  \ . \ee 
 These conditions determine the relative positions of the the black holes which do not preserve the same 4 supercharges at their respective horizons. The explicit form of $\omega$ can then be computed as in the case of $\N=2$ supergravity to be \cite{Denef}
 \begin{multline}  \omega =  i \sum_{\pA \ne \pB} \ \frac{ Q^{\pA\, IJ} Q^\pB_{IJ} - Q^\pA_{IJ} Q^{\pB\, IJ} }{|Êx_\pA - x_\pB| |Êx - x_\pA | |Êx - x_\pB| } \, \cdot \\*  \frac{ |Êx - x_\pA| - | x - x_\pB| - | x_\pA - x_\pB | }{ | x - x_\pB| | x_\pA - x_\pB| + ( x - x_\pB ) \cdot ( x_\pA - x_\pB ) } \varepsilon_{\mu\nu\sigma} ( x_\pA^\mu - x_\pB^\mu ) ( x^\nu - x_\pB^\nu ) d x^\sigma \ . \end{multline} 
Each term in the sum is a smooth $1$-form over $\IR^3$, and in particular, $\omega$ is regular in the neighbourhood of each horizon such that the black holes do not carry intrinsic angular momentum.\footnote{The local notion of angular momentum is hardly definable in general relativity, and what we mean is that the black holes do not carry ergospheres disjoint from their horizons, and that the induced metric on each horizon is spherically symmetric.}  However, space-time does carry a non-zero angular momentum 
\be J_{\mu\nu} =  \frac{i}{2}  \varepsilon_{\mu\nu\sigma} \, \sum_{\pA < \pB} \  \Scal{ Q^\pA_{IJ} Q^{\pB\, IJ} - Q^{\pA\, IJ} Q^{\pB}_{IJ} } \frac{ x_\pA^\sigma - x_\pB^\sigma }{ |x_\pA - x_\pB |} \ . \ee

\section{non-BPS black holes with saturated charges}

By truncation, these solutions reduce to general 1/4 BPS composites in $\N=4$ supergravity \cite{Ortin}, or 1/2 BPS composites in $\N=2$ supergravity \cite{Denef,Bates}. One can as well obtain non-BPS solutions as long as they correspond to charge configurations with a strictly positive quartic invariant, which is equivalent to the property that the Bogomolny bound is saturated,
\be M_{\scriptscriptstyle \rm ADM} = \max_\n[ \,   |Z_\n| \,  ]Ê\ . \ee
 This is for example a property of the non-BPS black holes with a vanishing central charge at the horizon in $\N=2,\, 4$ supergravity. We will now explain how such solutions can be obtained in the exceptional $\N=2$ supergravity theory. 

The exceptional supergravity theory admits scalar fields parameterising the symmetric space \be \cM_4 \cong \scal{U(1) \times E_{6(-78)} } \backslash  E_{7(-25)} \  ,
 \qquad \cM_3^*  \cong \scal{SL(2,\IR) \times E_{7(-25)}  } \backslash  E_{8(-24)}\ .\ee
 in four and three dimensions, respectively  \cite{GunaydinMagic}. The electromagnetic charges transform in the $\bf 56$ of $E_{7(-25)}$, which decomposes as $\IC \oplus {\bf 27}$ with respect to $U(1) \times E_{6(-78)}$, with ${\bf 27}$ being the complex fundamental representation of $ E_{6(-78)}$.
 
Before to consider the system of first order differential equations, let us fix our conventions by relating the special K\"{a}hler and the symmetric geometry of the moduli space $\cM_4$. We will represent the scalar fields by both their $E_{7(-25)}$ representative $v$ and the projective coordinates $z^i$ associated to the special K\"{a}hler geometry. The relevant quantities will be the central charge $Z$ and the matter charges $Z_a$,  which are defined as \cite{Ceresole:1995ca}
\bea  Z &=& v(q,p) = e^{\frac{\Ka}{2}} \Scal{Êq_\zero + z^i q_i + \sfrac{1}{2} c_{ijk} z^i z^j p^k - \sfrac{1}{6} c_{ijk} z^i z^j z^k p^\zero } \ , \CR
  Z_a &=& v(q,p)_a = \frac{1}{\sqrt{2}} V_a^{i} \, \scal{Ê \partial_i + \tfrac{1}{2} \partial_i \mathcal{K} }   Z  Ê\, , \eea
where $V_a^i$ and its complex conjugate $V^{\bar \imath\, a}$ are the inverse vielbeins 
\be V^i_a V^a_j  = \delta^i_j \ , \qquad V^{\bar \imath\, a} V_{\bar \jmath\, a} = \delta^{\bar \imath}_{\bar \jmath}Ê\ , \qquad V_i^a V_{\bar \imath\, a}Ê= \partial_i \partial_{\bar \imath} \mathcal{K} \ , \ee
and $\mathcal{K} $ is the K\"{a}hler potential defined in function of the $E_{6(-26)}$ invariant $c_{ijk}$ as \cite{GunaydinGeom}
\be \mathcal{K} = - \ln \Scal{Ê \, \frac{i}{6} \, c_{ijk} ( z^i - z^{\bar \imath} ) ( z^j - z^{\bar \jmath} ) ( z^k - z^{\bar k} )  }Ê  \ . \ee
The K\"{a}hler derivative and the corresponding K\"{a}hler spin-connexion are defined on the charges as 
\bea D Z &\equiv& d Z + \frac{1}{2} \scal{Ê \partial_i \mathcal{K} d z^i - \partial_{\bar \imath}Ê \mathcal{K} d z^{\bar \imath} } Z \ , \CR
 D Z_a &\equiv& d Z_a + \frac{1}{2} \scal{Ê \partial_i \mathcal{K} d z^i - \partial_{\bar \imath}Ê \mathcal{K} d z^{\bar \imath} } Z_a +  \scal{ÊV_{\bar \jmath \, a} \partial_i V^{\bar \jmath \, b } dz^i - V_j^b \partial_{\bar \imath}  V^j_a  d z^{\bar \imath} } \, Z_b \ . \label{CoKa} \eea 
The vielbeins $V_i^a$ are constrained such that these derivatives are covariant with respect to $U(1)\times E_{6(-78)}$. They satisfy the $E_{7(-25)}$ Bianchi identities
\be  D Z = 2 V^a Z_a \ ,\qquad D Z_a = V_a Z + 2 t_{abc} V^b Z^c \ , \label{BianM}Ê\ee
where $V_a$ is the coset component of the Maurer--Cartan form $d v v^{-1}$, 
\be  V_a =  \frac{1}{\sqrt{2}} V_{\bar \imath \, a} d z^{\bar \imath}   \ , \ee
and $t_{abc}$ is the symmetric invariant tensor of $E_{6(-78)} $, normalised such that the `adjoint identity' \cite{GunaydinGeom} reads
\be 3 \, t_{g\{ab} t_{cd\}h} t^{ghe}  = 2  \delta_{\{ a}{}^{\hspace{-1.8mm}e} \,  t_{bcd\}}  \ . \label{AdjI} \ee
The consistency between the Bianchi identities (\ref{BianM}) and the special geometry identities \cite{Strominger} implies that the vielbeins $V_i^a$ relate the cubic $E_{6(-26)}$ invariant $c_{ijk}$ to the cubic $E_{6(-78)}$ invariant $t_{abc}$ through 
\be t_{abc} = \frac{i}{\sqrt{2}} e^{ \Ka}   c_{ijk} V^i_a V^j_b V^k_c \ . \ee
The vielbeins $V_i^a$ are determined from this condition and the K\"{a}hler potential up to a $E_{6(-78)}$ local left multiplication. The electromagnetic fields $\xi^\Lambda,\, \tilde \xi_\Lambda$ are associated to the charges $q_\Lambda \Hat{=} ( q_\zero , q_i ) $ and $p^\Lambda \Hat{=} ( p^\zero , p^i )$, respectively. We define $v^\tt(d \xi)$ and $v^\tt(d\xi)_a$ such that 
\be q_\Lambda d \xi^\Lambda + p^\Lambda d \tilde{\xi}_\Lambda = \bar Z v^\tt(d \xi)  + 2 Z^a v^\tt(d \xi)_a + 2 Z_a v^\tt(d \xi)^a + Z \overline{ v^\tt(d \xi)}  \ . \ee

The non-BPS extremal solutions with a vanishing central charge at the horizon, $Z_*= 0$, are characterised by a Noether charge $\C_{|\Sigma}$ in the nilpotent orbit of $E_{8(-24)}$ which intersects the coset component $\e_{8(-24)} \ominus ( \sl_2 \oplus \e_{7(-25)} ) $ on the Lagrangian
\be \frac{SL(2,\IR) \times  E_{7(-25)} }{\scal{SO(2) \times Spin(10)} \ltimes \scal{ ({\bf 2}\otimes {\bf 16})^\ord{2} \oplus {\bf 1}^\ord{4} } } \subset \frac{E_{8(-24)}}{ E_{6(-14)} \ltimes \scal{{\bf 27}^\ord{2} \oplus {\bf 1}^\ord{4} } \times \IR } \, , \ee
in which lies the coset $1$-form $\bf P$. The generator ${\bf h} \in \e_{7(-25)} $ associated to the nilpotent representative $\bf P$, defines the following three-gradded decomposition of $\e_{7(-25)}$,
\be \e_{7(-25)} \cong {\bf 1}^\ord{-4} \oplus {\bf 32}^\ord{-2}\oplus \scal{ \gl_1 \oplus \so(2,10) }^\ord{0} \oplus {\bf 32}^\ord{2} \oplus {\bf 1}^\ord{4} \ . \ee
${\bf h}$ is a non-compact generator in the $\bf 27$ of the maximal compact subgroup $U(1) \times E_{6(-78)} \subset E_{7(-25)}$ which coefficients $\Omega_a$ satisfie 
\be  t^{abc} \Omega_b \Omega_c = 0 \ ,  \qquad \Omega^a \Omega_a = 2 \ .  \ee
These equations generalise (\ref{BPSattra}), and the first corresponds as well to the constraint satisfied by the charge $Z_{*\, a}$ at the horizon. Using the $E_{6(-78)}$ identity (\ref{AdjI}), one shows that $\Omega_a$ defines the three projectors 
\be \frac{1}{2}  \Omega_a \Omega^b   \ , \qquad \delta_a^b -  \frac{1}{2}  \Omega_a \Omega^b -  t_{ace}  t^{bde} \Omega^c \Omega_d  \ , \qquad t_{ace}  t^{bde} \Omega^c \Omega_d \ , \ee
associated to the decomposition of the ${\bf 27}$ of $E_{6(-78)}$ into the representations $\mathbb{C}^{\ord{4}} \oplus {\bf 16}^\ord{1} \oplus {\bf 10}^\ord{-2} $ of the stabiliser subgroup  $U(1) \times Spin(10) \subset E_{6(-78)}$ of $\Omega_a$. 

The first order system of differential equations associated to the extremal solutions of this type is then defined by $[{\bf h} , {\bf P } ]= 2 {\bf P}$, which reads
\bea d U + \frac{i}{2} e^{2U} \star d \omega &=& \Omega^a  e^{-U} v^\tt(d \xi)_a \ ,\CR
e^{-U} v^\tt(d \xi)_a &=& \frac{1}{2} \Omega_a \Scal{ d U +  \frac{i}{2} \star d \omega } + t_{abc} \Omega^b V^c \ , \hspace{15mm}e^{-U} v^\tt(d \xi) =\Omega^a V_a \ , \CR
V_a &=& \frac{1}{2} \Omega_a e^{-U} v^\tt(d \xi)  + t_{abc} \Omega^b e^{-U} v^\tt(d \xi)^c \ ,
\eea
with $\Omega_a$ satisfying the equation 
\be D \Omega_a - \frac{i}{2} e^{2U}  \star d \omega \, \Omega_a = 0 \label{E6Kill} \ee
where $D$ is the covariant derivative (\ref{CoKa}).

The integrability condition $[Ê{\bf h}Ê, [Ê{\bf h} , {\bf P} ]Ê]Ê= 4 {\bf P} $ gives 
\bea v^\tt(d\xi)_a &=& \frac{1}{2} \Omega_a \Omega^b\,  v^\tt(d\xi)_b  +  t_{ace}  t^{bde} \Omega^c \Omega_d \, v^\tt(d\xi)_b \ , \CR
V_a &=& \frac{1}{2} \Omega_a \Omega^b \,  V_b  +  t_{ace}  t^{bde} \Omega^c \Omega_d\, V_b \ , 
 \eea
 which is the condition that the component in the spinor representation of $U(1) \times Spin(10) $ vanishes for both $v^\tt(d\xi)_a$ and $V_a$. As for maximal supergravity, one concludes that the scalar fields take value in a subgroup $SL(2,\IR) \times_{\mathds{Z}_2} Spin(2,10) \subset  E_{7(-25)}$ and that only the phase of $\Omega_a$ varies over space-time. Such solutions are therefore 1/2 BPS solutions of the $\N=2$ supergravity which bosonic sector is defined by scalar fields parameterising the symmetric space $SO(2) \backslash SL(2,\IR) \times ( SO(2) \times SO(10) ) \backslash SO(2,10)$, and 12 electromagnetic fields in the vector representation of $SO(2,10)$. 

With the same reasoning as in the previous section, one obtains from the Ansatz 
 \be \xi^\Lambda = e^U \frac{\partial ( \Omega^a Z_a + \Omega_a Z^a)  }{ \partial q_\Lambda }  \ , \qquad  \tilde{\xi}_\Lambda = e^U \frac{\partial ( \Omega^a Z_a + \Omega_a Z^a ) }{ \partial p^\Lambda } \ ,\ee
that 
\be \Omega_a = 2 e^{U} v(\Ha)_a \    ,\ee
for harmonic functions $\Ha_\Lambda,\, \Ha^\Lambda$ which admit $E_{6(-14)} \subset E_{7(-25)}$ as a stabiliser subgroup, such that the scalar fields lie in their attractor manifold, or equivalently
\be t^{abc} v(\Ha)_b  v(\Ha)_c = 0 \ , \qquad v(\Ha) = 0 \, . \ee
Provided that the harmonic functions and the scalar fields moreover satisfy the constraint that 
\be v(d\Ha)_a = \frac{1}{2} \Omega_a \Omega^b v(d\Ha)_b  +  t_{ace}  t^{bde} \Omega^c \Omega_d v(d\Ha)_b \ , \ee
which is the statement that the component of $v(d\Ha)_a$ in the spinor representation of the $U(1) \times Spin(10) \subset E_{6(-78)}$ stabiliser subgroup of $v(\Ha)_a$ vanishes; one has the solution 
\bea e^{-2U} &=& 2 v(\Ha)^a v(\Ha)_a = \sqrt{ \lozenge(\Ha)} \ ,  \CR
 \xi^\Lambda  &=& \frac{1}{2 \lozenge(\Ha)} \frac{ \partial \lozenge(\Ha)}{ \partial \Ha_\Lambda}  \ , \qquad  \tilde{\xi}_\Lambda  = \frac{1}{2 \lozenge(\Ha) } \frac{ \partial \lozenge(\Ha) }{ \partial \Ha^\Lambda}   \ , \CR
 \frac{ \partial  \,  \Omega^a  Z_a(q,p) }{\partial q^\Lambda}  &=& \frac{1}{2 \sqrt[4]{\lozenge(\Ha)}} \Scal{Ê\frac{ \partial \sqrt{ \lozenge(\Ha)}}{ \partial \Ha_\Lambda} + i \Ha^\Lambda }  \ . \label{SoluM} \eea
where the scalar fields are only determined up to $32$ functions parameterising the attractor manifold 
\be \cM_{\hspace{-1mm} \mbox{ \tiny non-BPS} }  \cong \scal{ÊU(1) \times Spin(10) }Ê \backslash E_{6(-14)} \, , \ee
which are themselves determined in function of the asymptotic values of the scalar fields by the condition
\be V_ a = - e^{2U}   v(\Ha)_a \, v(d\Ha) - 2 e^{2U} \, t_{abc} \, v(\Ha)^b v(d\Ha)^c \ . \ee 
The Kaluza--Klein vector is determined as in the BPS case by 
\be d \omega =  \star \Scal{  \Ha^{\Lambda} d \Ha_{\Lambda} - \Ha_{\Lambda} d \Ha^{\Lambda} } \  . \ee
In order to obtain the explicit solution in the projective coordinates, it is convenient to decompose $z^i$ according to the irreducible representations $\mathds{R}^\ord{4} \oplus {\bf 16}^\ord{1} \oplus {\bf 10}^\ord{-2}$ of the subgroup $GL(1,\IR) \times Spin(1,9) \subset E_{6(-26)}$ associated to the truncation. Without lost of generality, we consider that the coordinates in the spinor representation vanish identically, such that only a scalar $z$ parameterising $SO(2) \backslash SL(2,\IR)$ and an $SO(1,9)$ vector $z^I$ parameterising $( SO(2)\times SO(10) ) \backslash SO(2,10)$  are non-zero. The K\"{a}hler potential then reduces to \footnote{The argument being positively defined on the domain covered by the coordinates $z,\, z^I$ \cite{GunaydinGeom}, the absolute value is redundant. However, it is required for equation (\ref{PT}) to be correct.} 
\be  \mathcal{K} = - \ln \Bigl|   \frac{i}{2}   \eta_{IJ} ( z - \bar z) ( z^I - \bar z^{I} ) ( z^J - \bar z^{J} )  \Bigr| Ê  \ , \ee 
where $\eta_{IJ}$ is the $SO(1,9)$ metric, and 
\be Z(z,z^I) = e^{\frac{\mathcal{K}}{2}} \Scal{Êq_\zero + z q_\un + z^I q_I + \sfrac{1}{2} \eta_{IJ}Êz^I z^J p^\un +  \eta_{IJ} z z^I p^J  - \sfrac{1}{2}Ê\eta_{IJ} zÊz^I z^J  p^\zero } \, .  \ee
Using this formula, one computes that 
\be \Omega_\asym^a  \, Z_a(z,z^I)  =  - (z - \bar z ) \, e^{-\frac{\mathcal{K}}{2}} \, \frac{\partial \, }{\partial z} \, \Scal{Êe^{\frac{\mathcal{K}}{2}} Z } = Z(\bar z , z^I) \, . \label{PT}  \ee
where $\Omega_{\asym\, a}$ is the asymptotic value of $\Omega_a$ which defines the truncation. Substituting the expression of $\Omega^a Z_a = e^{-i\alpha}  \Omega_\asym^a  \, Z_a(z,z^I) $ in (\ref{SoluM}), one obtains that 
\be z = \frac{ \frac{ \partial \sqrt{\lozenge(\Ha)} }{\partial \Ha_\un }  - i \Ha^\un }{ \frac{ \partial \sqrt{\lozenge(\Ha)} }{\partial \Ha_\zero }  - i \Ha^\zero } \ , \qquad z^I = \frac{ \frac{ \partial \sqrt{\lozenge(\Ha)} }{\partial \Ha_I }  + i \Ha^I }{ \frac{ \partial \sqrt{\lozenge(\Ha)} }{\partial \Ha_\zero } + i \Ha^\zero } \ , \label{ScalSol} \ee
such that the solution is the same as in the BPS case \cite{Bates,FerraraMagic}, up to the substitution of  $\bar z^I$ to $z^I$. The scalar fields are ensured to lie in the appropriate domain defining $\cM_4$ from the constraint that the harmonic functions admit $E_{6(-14)}$ as stabiliser subgroup of $E_{7(-25)}$. 

This property can be understood as follows. The involution $\iota$ defined by the $PT$ transformation in $S(Pin(1,2) \times O(2,10)) $,
\be \iota(q_\zero, \, q_\un , \, q_I ) = (q_\zero, -  q_\un , \, q_I )   \ ,  \quad \iota(p^\zero, \, p^\un , \, p^I ) = ( - p^\zero , \, p^\un , - p^I ) \ , \ee
defines an isomorphism between inequivalent orbits of $SL(2,\IR) \times SO(2,10)$ of stabiliser $SO(2) \times SO(10)$, which preserves the quartic invariant $\lozenge(q,p)$.  For trivial values of the scalar fields corresponding to the $STU$ truncation with $S = T = U = i$, one computes that if  $Z_a(q,p) = 0$, then $Z(\iota(q),\iota(p)) = 0$ and $Z_a(\iota(q),\iota(p)) = \frac{1}{2}ÊZ(q,p) \Omega_{\asym \, a}$, such that in general, this involution maps charges of stabiliser subgroup  $E_{6(-78)}$ to charges of stabiliser subgroup $E_{6(-14)}$ in $E_{7(-25)}$. This involution extends to an involution acting on the solutions as
\begin{gather} \iota(\Ha_\zero, \, \Ha_\un , \, \Ha_I ) = (\Ha_\zero, -  \Ha_\un , \, \Ha_I )   \ ,  \quad \iota(\Ha^\zero, \, \Ha^\un , \, \Ha^I ) = ( - \Ha^\zero , \, \Ha^\un , - \Ha^I ) \ , \CR
\iota_*U = U \ , \qquad \iota_*\omega = \omega \ , \qquad \iota_* \xi^\Lambda = \xi^\Lambda \ , \qquad \iota_* \tilde{\xi}_\Lambda = \tilde{\xi}_\Lambda  \ , \CR
 \iota_* z    = z    \ , \qquad  \iota_* z^I  =  \bar z^I  \ , 
\end{gather}
which preserves the BPS first order system of differential equations, although does not preserve the BPS condition itself, similarly as in \cite{Galli}. One can then check using (\ref{ScalSol}) that
\be \iota_* z(\iota(\Ha)) =  - \bar z(\Ha) \ , \qquad \iota_* z^I(\iota(\Ha)) = z^I(\Ha) \ , \ee
such that the domain of definition of $z^I$ is obviously preserved, and the imaginary part of $z$ remains strictly positive. 

This involution acts on the Noether charge ${\bf Q}_{|\Sigma}$ as a $PT$ transformation in $S( O(2,2) \times O(2,10) ) \subset SO(4,12)$, 
\be \iota(M, \, 0 ,\, q_\zero, \, q_\un , \, q_I ,\, p^\zero, \, p^\un , \, p^I ,\, \Sigma ,\, \Sigma^I ) = ( M, - 0,\, q_\zero,  - q_\un , \, q_I , -p^\zero, \, p^\un , - p^I ,- \bar \Sigma ,\, \Sigma^I) \ . \ee
The action of the $\mathds{Z}_2$ centre of $SO(4,12)$ on $\so(4,12)$ indeed defines an isomorphism between the two inequivalent nilpotent orbits of the connected component of $SO(4,12)$ of reducible stabiliser $SO(2) \times SO(10)$ \cite{DokovicSO} corresponding to the BPS and the non-BPS solutions with $Z_* = Z^I_* = 0$, respectively, as discussed in \cite{BossardW} in the $STU$ model. 

Note finally that 
\be \iota_*U(\iota(\Ha)) = U(\Ha) \ , \qquad \iota_* \omega(\iota(\Ha)) = - \omega(\Ha) \ , \ee
such that the involution acts on the space-time geometry as a parity transformation. 

\pagebreak[4]

\noindent
{\bf Acknowledgments}: I would like to thank Dumitru Astefanesei, Iosif Bena, Stefano Giusto, Hermann Nicolai, Boris Pioline, Maria J.~Rodriguez and Cl\'ement Ruef for stimulating discussions.



\begin{thebibliography}{99}


\bibitem{PiolineLecture} 
B.~Pioline,
 ``Lectures on on black holes, topological strings and quantum attractors,'' 
  Class.\ Quant.\ Grav.\  {\bf 23}  (2006) S981 
  \eprint{hep-th/0607227}.

\bibitem{Denef}
  F.~Denef,
  ``Supergravity flows and $D$-brane stability,''
  JHEP {\bf 0008} (2000) 050
  \eprint{hep-th/0005049}.


\bibitem{Bates}
  B.~Bates and F.~Denef,
  ``Exact solutions for supersymmetric stationary black hole composites,''
  \eprint{hep-th/0304094}.


\bibitem{VVD}
  R.~Dijkgraaf, E.~P.~Verlinde and H.~L.~Verlinde,
  ``Counting dyons in $\N=4$ string theory,''
  Nucl.\ Phys.\  B {\bf 484} (1997) 543
  \eprint{hep-th/9607026}.

\bibitem{StroVD}
  D.~Shih, A.~Strominger and X.~Yin,
  ``Recounting dyons in $\N = 4$ string theory,''
  JHEP {\bf 0610} (2006) 087
  \eprint{hep-th/0505094}.

\bibitem{JatkarSen}
  J.~R.~David, D.~P.~Jatkar and A.~Sen,
  ``Product representation of dyon partition function in CHL models,''
  JHEP {\bf 0606} (2006) 064
  \eprint{hep-th/0602254}.

\bibitem{Dabholkar}
  A.~Dabholkar and S.~Nampuri,
  ``Spectrum of dyons and black holes in CHL orbifolds using Borcherds lift,''
  JHEP {\bf 0711} (2007) 077
  \eprint{hep-th/0603066}.


\bibitem{Dimitru}
  D.~Astefanesei, K.~Goldstein and S.~Mahapatra,
  ``Moduli and (un)attractor black hole thermodynamics,''
  Gen.\ Rel.\ Grav.\  {\bf 40} (2008) 2069
  \eprint{hep-th/0611140}.


\bibitem{DabhoSen}
  A.~Dabholkar, A.~Sen and S.~P.~Trivedi,
  ``Black hole microstates and attractor without supersymmetry,''
  JHEP {\bf 0701} (2007) 096
  \eprint{hep-th/0611143}.


\bibitem{FerraraAttra1}
  S.~Ferrara, R.~Kallosh and A.~Strominger,
  ``$\N=2$ extremal black holes,''
  Phys.\ Rev.\  D {\bf 52} (1995) 5412
  \eprint{hep-th/9508072}.

\bibitem{FerraraAttra2}
  S.~Ferrara and R.~Kallosh,
  ``Universality of supersymmetric attractors,''
  Phys.\ Rev.\  D {\bf 54} (1996) 1525
  \eprint{hep-th/9603090}.

\bibitem{FerraraAttra3}
  S.~Ferrara, G.~W.~Gibbons and R.~Kallosh,
  ``Black holes and critical points in moduli space,''
  Nucl.\ Phys.\  B {\bf 500} (1997) 75
  \eprint{hep-th/9702103}.

\bibitem{SenAttra}
  A.~Sen,
  ``Black hole entropy function and the attractor mechanism in higher
  derivative gravity,''
  JHEP {\bf 0509} (2005) 038
  \eprint{hep-th/0506177}.
  
  
\bibitem{GoldsteinAttra}
  K.~Goldstein, N.~Iizuka, R.~P.~Jena and S.~P.~Trivedi,
  ``Non-supersymmetric attractors,''
  Phys.\ Rev.\  D {\bf 72} (2005) 124021
  \eprint{hep-th/0507096}.

\bibitem{Ortin}
  J.~Bellorin and T.~Ortin,
  ``All the supersymmetric configurations of $\N = 4$, $d = 4$ supergravity,''
  Nucl.\ Phys.\  B {\bf 726} (2005) 171 
  \eprint{hep-th/0506056}.

\bibitem{FerraraMagic}
  S.~Ferrara, E.~G.~Gimon and R.~Kallosh,
  ``Magic supergravities, $\N = 8$ and black hole composites,''
  Phys.\ Rev.\  D {\bf 74}  (2006) 125018
  \eprint{hep-th/0606211}.

\bibitem{Ceresole:2007wx}
  A.~Ceresole and G.~Dall'Agata,
  ``Flow equations for non-BPS extremal black holes,''
  JHEP {\bf 0703}, 110 (2007)
  \eprint{hep-th/0702088}.

\bibitem{Andrianopoli:2007gt}
  L.~Andrianopoli, R.~D'Auria, E.~Orazi and M.~Trigiante,
  ``First order description of black holes in moduli space,''
  JHEP {\bf 0711} (2007) 032
  \eprintN{0706.0712}.

\bibitem{Lopes Cardoso:2007ky}
  G.~Lopes Cardoso, A.~Ceresole, G.~Dall'Agata, J.~M.~Oberreuter and J.~Perz,
  ``First-order flow equations for extremal black holes in very special
  geometry,''
  JHEP {\bf 0710} (2007) 063
  \eprintN{0706.3373}.

\bibitem{Perz:2008kh}
  J.~Perz, P.~Smyth, T.~Van Riet and B.~Vercnocke,
  ``First-order flow equations for extremal and non-extremal black holes,''
  JHEP {\bf 0903} (2009) 150
  \eprintN{0810.1528}.


\bibitem{Hotta:2007wz}
  K.~Hotta and T.~Kubota,
  ``Exact solutions and the attractor mechanism in non-BPS black holes,''
  Prog.\ Theor.\ Phys.\  {\bf 118} (2007) 969
  \eprintN{0707.4554}.


\bibitem{Gimon:2007mh}
  E.~G.~Gimon, F.~Larsen and J.~Simon,
  ``Black holes in supergravity: the non-BPS branch,''
  JHEP {\bf 0801} (2008) 040
  \eprintN{0710.4967}.

\bibitem{Bellucci:2008sv}
  S.~Bellucci, S.~Ferrara, A.~Marrani and A.~Yeranyan,
  ``$stu$ black holes unveiled,''
  \eprintN{0807.3503}.
  
\bibitem{Gimon:2009gk}
  E.~G.~Gimon, F.~Larsen and J.~Simon,
  ``Constituent Model of Extremal non-BPS Black Holes,''
  JHEP {\bf 0907} (2009) 052
  \eprintN{0903.0719}.

\bibitem{Gaiotto:2007ag}
  D.~Gaiotto, W.~W.~Li and M.~Padi,
  ``Non-supersymmetric attractor flow in symmetric spaces,''
  JHEP {\bf 0712} (2007) 093
  \eprintN{0710.1638}.

\bibitem{Ceresole:2009iy}
  A.~Ceresole, G.~Dall'Agata, S.~Ferrara and A.~Yeranyan,
  ``First order flows for $\N=2$ extremal black holes and duality invariants,''
  Nucl.\ Phys.\  B {\bf 824} (2010) 239
  \eprintN{0908.1110}.



\bibitem{Ferrara:2009bw}
  S.~Ferrara, A.~Marrani and E.~Orazi,
  ``Maurer--Cartan equations and black hole superpotentials in $\N =8$
  supergravity,''
  \eprintN{0911.0135}.

\bibitem{BossardW}
  G.~Bossard, Y.~Michel and B.~Pioline,
  ``Extremal black holes, nilpotent orbits and the true fake superpotential,''
  \eprintN{0908.1742}.

\bibitem{Ceresole:2009vp}
  A.~Ceresole, G.~Dall'Agata, S.~Ferrara and A.~Yeranyan,
  ``Universality of the superpotential for $d = 4$ extremal black holes,''
  \eprintN{0910.2697}.


\bibitem{NicolaiPM}
  G.~Bossard and H.~Nicolai,
  ``Multi-black holes from nilpotent Lie algebra orbits,''
  \eprintN{0906.1987}.

\bibitem{BossardPM}
  G.~Bossard,
  ``The extremal black holes of $\N=4$ supergravity from $\so(8,2+n)$ nilpotent
  orbits,''
  \eprintN{0906.1988}.
  
  \bibitem{Bena1}
  I.~Bena, S.~Giusto, C.~Ruef and N.~P.~Warner,
  ``Multi-center non-BPS black holes - the solution,''
  JHEP {\bf 0911}  (2009) 032
  \eprintN{0908.2121}.

\bibitem{Bena2}
  I.~Bena, S.~Giusto, C.~Ruef and N.~P.~Warner,
  ``Supergravity solutions from floating branes,''
  \eprintN{0910.1860}.


\bibitem{Galli}
  P.~Galli and J.~Perz,
  ``Non-supersymmetric extremal multicenter black holes with superpotentials,''
  \eprintN{0909.5185}.


\bibitem{Collingwood}
D.~Collingwood and W.~ McGovern,
``Nilpotent orbits in semisimple Lie algebras"
Van Nostrand Reinhold Mathematics Series,  New York, 1993.

\bibitem{Gunaydin:2005mx}
  M.~G\"unaydin, A.~Neitzke, B.~Pioline and A.~Waldron,
  ``BPS black holes, quantum attractor flows and automorphic forms,''
  Phys.\ Rev.\  D {\bf 73} (2006) 084019
  \eprint{hep-th/0512296}.

\bibitem{BossardNil}
  G.~Bossard, H.~Nicolai and K.~S.~Stelle,
  ``Universal BPS structure of stationary supergravity solutions,''
   JHEP {\bf 0907} (2009) 003 
  \eprintN{0902.4438}.

\bibitem{CremmerJulia}
  E.~Cremmer and B.~Julia,
  ``The $SO(8)$ Supergravity,''
  Nucl.\ Phys.\  B {\bf 159} (1979) 141.

\bibitem{de Wit:1982ig}
  B.~de Wit and H.~Nicolai,
  ``$\N=8$ Supergravity,''
  Nucl.\ Phys.\  B {\bf 208} (1982) 323.


\bibitem{Breitenlohner:1987dg}
  P.~Breitenlohner, D.~Maison and G.~W.~Gibbons,
  ``Four-dimensional black holes from Kaluza--Klein theories,''
  Commun.\ Math.\ Phys.\  {\bf 120} (1988) 295.


\bibitem{BossardNUT}
  G.~Bossard, H.~Nicolai and K.~S.~Stelle,
  ``Gravitational multi-NUT solitons, Komar masses and charges,''
  Gen.\ Rel.\ Grav.\  {\bf 41} (2009)1367 
  \eprintN{0809.5218}.


  \bibitem{Ferrara:1997uz}
  S.~Ferrara and M.~G\"unaydin,
  ``Orbits of exceptional groups, duality and BPS states in string theory,''
  Int.\ J.\ Mod.\ Phys.\  A {\bf 13} (1998) 2075
  \eprint{hep-th/9708025}.

\bibitem{Bellucci:2006xz}
  S.~Bellucci, S.~Ferrara, M.~G\"unaydin and A.~Marrani,
  ``Charge orbits of symmetric special geometries and attractors,''
  Int.\ J.\ Mod.\ Phys.\  A {\bf 21} (2006) 5043
  \eprint{hep-th/0606209}.

\bibitem{GunaydinMagic}
  M.~G\"unaydin, G.~Sierra and P.~K.~Townsend,
  ``Exceptional supergravity theories and the magic square,''
  Phys.\ Lett.\  B {\bf 133} (1983) 72.
  

\bibitem{Andrianopoli:1997wi}
  L.~Andrianopoli, R.~D'Auria, S.~Ferrara, P.~Fre and M.~Trigiante,
  ``$E_{7(7)}$ duality, BPS black-hole evolution and fixed scalars,''
  Nucl.\ Phys.\  B {\bf 509} (1998) 463
  \eprint{hep-th/9707087}.
 
\bibitem{Ceresole:1995ca}
  A.~Ceresole, R.~D'Auria and S.~Ferrara,
  ``The symplectic structure of $\N=2$ supergravity and its central extension,''
  Nucl.\ Phys.\ Proc.\ Suppl.\  {\bf 46} (1996) 67
  \eprint{hep-th/9509160}.



\bibitem{GunaydinGeom}
  M.~Gunaydin, G.~Sierra and P.~K.~Townsend,
  ``The geometry of $\N=2$ Maxwell--Einstein supergravity and Jordan algebras,''
  Nucl.\ Phys.\  B {\bf 242} (1984) 244.


\bibitem{Strominger}
  A.~Strominger,
  ``Special geometry,''
  Commun.\ Math.\ Phys.\  {\bf 133} (1990) 163.


  
  \bibitem{DokovicSO}
 D.~\v{Z}.~\DJo, N. Lemire, J. Sekiguchi, 
``The closure ordering of adjoint nilpotent orbits in $\so(p,q)$,
Tohoku Mat.\ J.\ {\bf 53}  (2001) 395.




\end{thebibliography}
\end{document}